\documentclass[conference]{IEEEtran}
\IEEEoverridecommandlockouts
\usepackage[T1]{fontenc}
\usepackage[utf8]{inputenc}
\usepackage{amsmath,amssymb,mathtools}
\usepackage{graphicx}
\usepackage{booktabs}
\usepackage{array}
\usepackage{cite}
\usepackage{microtype}
\usepackage{url}
\usepackage{xcolor, soul}
\usepackage{balance}
\usepackage{changes}

\newcommand{\pos}[1]{\left[#1\right]^+}

\usepackage{float}
\usepackage{titlesec}

\titlespacing*{\section}
{0pt}{1.5ex plus 0.5ex minus 0.2ex}{0.8ex}

\titlespacing*{\subsection}
{0pt}{1.2ex plus 0.4ex minus 0.2ex}{0.5ex}

\titlespacing*{\subsubsection}
{0pt}{1ex plus 0.3ex minus 0.2ex}{0.4ex}

\def\BibTeX{{\rm B\kern-.05em{\sc i\kern-.025em b}\kern-.08em
    T\kern-.1667em\lower.7ex\hbox{E}\kern-.125emX}}

\title{Grid-Responsive Pricing of Dynamic Wireless Charging with Coupled Residential Demand}
\author{\IEEEauthorblockN{Zhiheng Wang, Dipanjan Ghose, S. Sivaranjani, Junjie Qin, and Qin Lv*
\thanks{This work is supported in part by the U.S. National Science Foundation under Grant No. EEC-1941524. ZW and QL are with the University of Colorado Boulder, Boulder, CO, USA. DG, SS, and JQ are with Purdue University, West Lafayette, IN, USA. \textit{*Correspondence:} qin.lv@colorado.edu}}}

\begin{document}
\maketitle

\begin{abstract}
Dynamic wireless charging (DWC) is an emerging technology that enables electric vehicles (EVs) to recharge while traveling. Its unique operational model creates two revenue streams for DWC operators through lane access tolls and electricity sales. 
However, these prices influence not only traffic flow and roadway energy delivery but also how much EV charging moves to homes instead, potentially within the same distribution network.
Neglecting this coupling can underestimate grid loading and compromise pricing decisions. We develop a coupled traffic, EV energy, driver-response, and AC distribution network model with a model predictive control (MPC) framework that jointly optimizes DWC tolls and charging prices while explicitly accounting for displaced residential charging and network losses. Using 78 days of I-210 West traffic data, we show that conventional MPC exceeds feeder loading limits on 52 days, whereas the proposed residential load aware controller eliminates these violations with only 0.6\% lower operator revenue. Under a more constrained feeder configuration, it eliminates 50 overloaded days with a 0.9\% revenue reduction. Sensitivity analyses highlight the importance of accurate residential demand estimation and feeder loss modeling, while peak-aware pricing further reduces daily peak loading by 3.7\% relative to the base load. These results demonstrate the value of coordinated pricing that captures the interaction between roadway and residential charging.
\end{abstract}

\section{Introduction}\label{sec:intro}

Dynamic wireless charging (DWC) enables electric vehicles (EVs) to receive energy while traveling over electrified roadways. Several pilot projects worldwide have demonstrated the technical feasibility of DWC and its potential to reduce charging downtime and vehicle queuing compared with stationary charging stations (SCS)~\cite{electreonprojects}. However, DWC infrastructure requires substantial capital investment~\cite{Ghose2026Rethinking}, making the economically viable operation of these systems an important consideration.

Unlike SCS, DWC infrastructure is integrated directly into the roadway, and its utilization depends on traffic movement and lane usage. Further, DWC provides a charging service to EVs traveling along the corridor, differentiating it from conventional toll roads. Thus, under a pricing structure that charges EVs for both access to the electrified lane and energy delivered, DWC operators can generate revenue through two distinct mechanisms: roadway tolls and electricity sales.

Determining toll and charging prices is critical, as tolls influence traffic demand~\cite{huan2019toll}, while electricity prices affect EV charging decisions. Together, these prices determine DWC lane utilization, traffic congestion, and distribution network demand. EVs receiving insufficient energy through DWC may later charge at home, potentially on the same feeders supplying the roadway. Thus, pricing decisions also influence residential charging and aggregate feeder loading. This raises an important question: \textit{How can DWC operators coordinate tolls and charging prices to balance revenue, traffic performance, and electrical impacts on shared distribution networks?}

Previous studies have investigated dynamic tolling for traffic management~\cite{yin2009,tan2018}, coordinated pricing in coupled transportation and electricity networks~\cite{alizadeh2017,zhou2022}, and pricing and charging control for DWC systems~\cite{liu2024toll,lai2025wcl}. However, these approaches do not capture how DWC pricing redistributes charging demand between electrified roadways and nearby residences. Residential charging is not represented as a demand component that responds to DWC utilization, potentially leading to inaccurate loading estimates on shared distribution feeders.

Here, we develop a grid-aware DWC pricing framework that explicitly couples roadway charging decisions with traffic dynamics and residual residential demand. Our main contributions are: (a) We develop a coupled transportation and power system model that captures how tolls and charging prices influence lane utilization, DWC energy delivery, and subsequent residential charging on shared feeders. (b) We formulate a receding-horizon joint pricing controller that incorporates reconstructed residential demand and AC feeder losses into feeder loading predictions. (c) Using 78 days of I-210 West traffic data, we demonstrate that accounting for this coupling eliminates observed overloading  with less than 1\% revenue loss, while examining sensitivity to demand estimation errors and trade-offs among revenue, traffic performance, and peak demand. Section~\ref{sec:system} presents the formulation, Section~\ref{sec:case} presents the case study and results, and Section~\ref{sec:conclusion} concludes the paper.

\section{Problem Formulation}\label{sec:system}

\subsection{System Overview and Notation}\label{sec:overview}

We consider an electrified freeway corridor where one existing lane is converted to DWC alongside general-purpose (GP) lanes. All traffic consists of EVs with heterogeneous energy requirements. The DWC lane shares distribution feeders with residential charging and existing loads, subject to an aggregate feeder-head apparent power limit $S^{\max}$ (MVA).

Let $\mathcal C$ and $\mathcal L\!=\!\{\mathrm{GP},\mathrm{DWC}\}$ denote EV cohorts and lane groups. We index simulation steps by $t\in\mathcal T$ and pricing intervals by $k\!\in\!\mathcal K\!=\!\{1,\ldots,K\}$, with durations $\Delta t$ and $\Delta_k$ (h). Here, $k(t)$ identifies the pricing interval at $t$, and $\mathcal T_k$ contains its simulation steps. Every interval, DWC operator sets entry toll $\tau_k$ (USD) and charging price adjustment (CPA) $\mu_k$ (USD/kWh).

\subsection{Traffic Dynamics and Lane Choice}\label{sec:traffic}

We model traffic propagation using the cell transmission model (CTM)~\cite{daganzo1994}, implemented through CTMSIM~\cite{kurzhanskiy2009}. We discretize each lane group $\ell\in\mathcal L$ into $I$ roadway cells, indexed by $\mathcal I=\{1,\ldots,I\}$, with lengths $\Delta x_i$ (km). Let $n_{i,t}^{\ell}$ denote the number of EVs occupying cell $i$ of lane $\ell$ at timestep $t$. We update cell populations using CTM sending and receiving flows, accounting for asymmetric node interactions, ramp capacity constraints, and lane transfers~\cite{daganzo1994}. Since DWC demand depends on lane utilization, EVs requesting DWC access are admitted subject to available lane capacity and density:
\begin{equation}\label{eq:lc}
n_{i,t}^{\mathrm{lc}}=\min\left\{
n_{i,t}^{\mathrm{want}},\,
\kappa\Delta t M_i^{\mathrm{DWC}},\,
\pos{\xi d_{i,t}-n_{i,t}^{\mathrm{arr}}}
\right\},
\end{equation}
where $[\cdot]^+\!=\!\max\{0,\cdot\}$, $n_{i,t}^{\mathrm{want}}$ denotes eligible EVs requesting entry, $M_i^{\mathrm{DWC}}$ is the flow capacity of DWC cell $i$ (veh/h), and $n_{i,t}^{\mathrm{arr}}$ denotes EVs arriving from the upstream DWC cell. The available capacity is $d_{i,t}\!=\!\rho_i^{\mathrm{crit}}\Delta x_i\!-\!n_{i,t}^{\mathrm{DWC}}$, where $n_{i,t}^{\mathrm{DWC}}$ counts EVs already in DWC cell $i$ and $\rho_i^{\mathrm{crit}}$ is the critical density (veh/km)~\cite{ghose2025dwptgrid}. Parameters $\kappa,\xi\!\in\![0,1]$ limit transfer capacity and usable lane capacity, respectively. We permit transfers only upstream of each EV's exit.

We group EVs into cohorts $c\!\in\!\mathcal C$ based on EV class, battery energy, exit location, lane and charging decisions. Let $n_{i,c,t}^{\ell}$ denote the number of cohort-$c$ EVs in cell $i$ of lane group $\ell$ at timestep $t$, such that $n_{i,t}^{\ell}=\sum_{c\in\mathcal C}n_{i,c,t}^{\ell}$. We distribute CTM flows across cohorts proportional to their population.

\subsection{EV Energy Dynamics and DWC Demand}\label{sec:energy}

Let $e_{i,c,t}$ denote the battery energy (kWh) per EV in cohort $c$ in cell $i$ at timestep $t$, accounting for driving and auxiliary energy consumption \cite{ghose2025dwptgrid}. Each EV in cohort $c$ has battery capacity $e_c^{\mathrm{cap}}$ (kWh) and target energy $e_c^{\mathrm{tgt}}$ (kWh). For each EV of cohort $c$ charging in DWC cell $i$, the energy delivered at $t$ is:
\noindent\begin{equation}\label{eq:charge}
\Delta e_{i,c,t}\!=\!\min\!\left\{
p^{\mathrm{veh}}\Delta t,
\pos{e_c^{\mathrm{tgt}}\!-\!e_{i,c,t}}\!,
\!\pos{e_c^{\mathrm{cap}}\!-\!e_{i,c,t}}\!
\right\},
\end{equation}
where $p^{\mathrm{veh}}$ (kW) is the maximum charging power per EV.

Let $n_{i,c,t}^{\mathrm{chg}}$ denote the number of charging EVs of cohort $c$ in DWC cell $i$. The aggregate delivered battery energy (kWh) is:
\begin{equation}\label{eq:etot}
E_t^{\mathrm{del}} =
\sum_{i\in\mathcal I}\sum_{c\in\mathcal C}
n_{i,c,t}^{\mathrm{chg}}\Delta e_{i,c,t}.
\end{equation}

We convert the total battery energy delivered to grid-side DWC power (MW) as $p_t^{\mathrm{DWC}}=10^{-3}E_t^{\mathrm{del}}/(\eta\Delta t)$, where $\eta\!=\!0.90$ is the charging efficiency. Summing this power over pricing intervals gives the cumulative grid energy drawn by DWC (kWh), $E_k^{\mathrm{DWC}}=10^3\Delta_k\sum_{k'=1}^{k}\bar p_{k'}^{\mathrm{DWC}}$, where $\bar p_{k'}^{\mathrm{DWC}}$ is the average DWC power during interval $k'$.

\subsection{Driver Response to Toll and Charging Prices}\label{sec:choice}

We model EV lane choice and charging participation as functions of $\tau_k$ and $\mu_k$. Let $\pi^{\mathrm{DWC}}$ and $\pi^{\mathrm{res}}$ denote DWC and residential electricity prices (USD per battery kWh). The effective DWC price is $\pi^{\mathrm{DWC}}+\mu_k$, yielding potential savings $\Delta\pi_k=\pos{\pi^{\mathrm{res}}-\pi^{\mathrm{DWC}}-\mu_k}$ per battery kWh.

As EVs enter the corridor, we group arrivals into cohorts $m$ by entry location, vehicle class, and exit. For each $m$ arriving during interval $k$, we model the EV fraction requesting DWC access, the fraction canceling trips, and charging probability:
\begin{align}
\phi_{m,k}^{\mathrm{DWC}}
&=\min\{1,\phi_0\exp[-\varepsilon(\tau_k-\Delta\pi_k e_m^{\mathrm{ref}})]\},
\label{eq:choice}\\
\phi_{m,k}^{\mathrm{skip}}
&=(1-\chi)\pos{\phi_0-\phi_{m,k}^{\mathrm{DWC}}},
\label{eq:suppress}\\
\phi_k^{\mathrm{chg}}
&=\min\{1,\exp(-\beta\mu_k)\},
\label{eq:optin}
\end{align}
where $e_m^{\mathrm{ref}}$ is the maximum available DWC energy (kWh) before exit, $\phi_0$ is the reference lane share when $\tau_k=\Delta\pi_k e_m^{\mathrm{ref}}$, $\varepsilon$ (USD$^{-1}$) and $\beta$ (kWh/USD) are price sensitivities, and $\chi$ is the fraction declining DWC access but continuing their trips. These assumed, uncalibrated responses govern lane requests in \eqref{eq:lc} and charging participation in \eqref{eq:etot}, while cohorts $c\in\mathcal C$ track EVs through the CTM.

\subsection{Residential Charging and Distribution Network Model}\label{sec:feeder}

DWC pricing influences residential charging by changing the energy EVs receive on the roadway. EVs receiving less DWC energy require more residential charging, potentially increasing loading on shared feeders. We assume EVs meet their remaining energy requirements at home.

Let $E^{\mathrm{res}}$ (kWh) denote the aggregate energy needed to reach the charging targets of EVs exiting or remaining on the corridor, including the corresponding charging requirements of canceled trips without DWC. A fraction $\sigma$ of this demand is served by the $N$ identical feeders supplying DWC. Given a normalized charging profile $h_k$ (h$^{-1}$), satisfying $\Delta_k\sum_{k\in\mathcal K}h_k=1$, residential charging power (MW) is $p_k^{\mathrm{res}}=10^{-3}\sigma E^{\mathrm{res}}h_k/\eta$. We calculate realized residential demand retrospectively because $E^{\mathrm{res}}$ depends on total daily DWC delivery. For $\sigma<1$, DWC introduces additional net energy demand on the shared feeders.

We model each feeder as an independent $J$-bus radial network with base active and reactive load vectors $\mathbf p_k^{\mathrm{base}},\mathbf q_k^{\mathrm{base}}\in\mathbb R^J$ (MW, Mvar). We distribute residential charging across buses proportional to their base active loads at unity power factor, while DWC connects at bus $j^{\mathrm{DWC}}$ with power factor $\cos\varphi$. Assuming constant-power loads, the resulting bus demands are:
\begin{align}
\mathbf p_k &=
\mathbf p_k^{\mathrm{base}}
+\frac{p_k^{\mathrm{res}}}{N}
\frac{\mathbf p_k^{\mathrm{base}}}
{\mathbf 1^\top\mathbf p_k^{\mathrm{base}}}
+\frac{\bar p_k^{\mathrm{DWC}}}{N}\mathbf 1_{j^{\mathrm{DWC}}},
\label{eq:busp}\\
\mathbf q_k &=
\mathbf q_k^{\mathrm{base}}
+\frac{\bar p_k^{\mathrm{DWC}}}{N}
\tan\varphi\,\mathbf 1_{j^{\mathrm{DWC}}},
\label{eq:busq}
\end{align}
where $\mathbf 1$ is the all-ones vector and $\mathbf 1_{j^{\mathrm{DWC}}}$ selects the DWC connection bus. 
We compute aggregate feeder loading as $S_k=N\mathcal F_{\mathrm{AC}}(\mathbf p_k,\mathbf q_k)$, where $\mathcal F_{\mathrm{AC}}$ returns feeder-head apparent power (MVA) from AC power flow~\cite{BaranWu1989DistFlow}, accounting for losses.

\subsection{Joint Toll and Charging Pricing Controller}\label{sec:control}

We develop a model predictive control (MPC) framework to jointly optimize DWC tolls and charging prices while balancing operator revenue, traffic performance, and feeder loading.

At the beginning of each interval $k$, we select the price vector $\mathbf u_k=(\tau_k,\mu_k)^\top\in\mathcal B$, where $\mathcal B$ denotes the admissible price bounds. We evaluate candidate prices over a receding horizon $\mathcal H_k=\mathcal T_k\cup\mathcal T_{k+1}\cup\mathcal T_{k+2}$ (45 min), holding each candidate constant throughout the horizon. The candidate set is $\mathcal U_k:=\{\mathbf u_{k-1}+(a_\tau\delta_\tau,a_\mu\delta_\mu)^\top: a_\tau,a_\mu\in\{-1,0,1\}\}\cap\mathcal B$,
where $\delta_\tau$ (USD) and $\delta_\mu$ (USD/kWh) are the toll and CPA increments.

For each candidate, we simulate traffic, EV energy, and driver responses over the prediction horizon using known demand and wholesale prices. Residential charging is estimated separately, as its realized demand depends on total daily DWC delivery. We define the economic and traffic performance term $g_t=w_RR_t-w_DD_t-w_Sn_t^{\mathrm{sup}}$ (USD) and the overload penalty $o_{t,k}=60\Delta t\pos{\hat S_t+\Delta S_k-S^{\max}}$. The controller solves:
\begin{equation}\label{eq:score}
\begin{aligned}
\min_{\mathbf u\in\mathcal U_k}\quad
&\sum_{t\in\mathcal H_k}(w_Lo_{t,k}-g_t)\\
\mathrm{s.t.}\quad
&\text{CTM dynamics and \eqref{eq:lc}--\eqref{eq:optin} over }\mathcal H_k.
\end{aligned}
\end{equation}

Here $R_t$ (USD) is operator net revenue, $D_t$ (veh$\cdot$h) is traffic delay, and $n_t^{\mathrm{sup}}$ (veh) counts canceled trips exceeding 1\% of potential arrivals at timestep $t$. The objective weights are $w_R$ (dimensionless), $w_D$ (USD/(veh$\cdot$h)), $w_S$ (USD/veh), and $w_L$ (USD/(MVA$\cdot$min)). The predicted feeder loading $\hat S_t$ (MVA) and correction $\Delta S_k$ (MVA) give $o_{t,k}$ (MVA$\cdot$min). Starting from $\mathbf u_0=\mathbf 0$ each day, the controller evaluates up to nine candidates and implements the minimum-cost decision each interval. The feeder loading limit is imposed through a soft penalty rather than a hard constraint.

\textit{1) Lane-only prediction:}
The baseline MPC predicts feeder loading from existing demand and DWC power, without accounting for residential charging:
\begin{equation}\label{eq:pred}
\hat S_t=
\sqrt{
(P_t^{\mathrm{base}}+p_t^{\mathrm{DWC}})^2+
(Q_t^{\mathrm{base}}+p_t^{\mathrm{DWC}}\tan\varphi)^2
},
\end{equation}
where $P_t^{\mathrm{base}}$ (MW) and $Q_t^{\mathrm{base}}$ (Mvar) are aggregate feeder-head active and reactive powers of the base loads, including their losses. This prediction excludes residential charging and incremental losses from charging demand, with $\Delta S_k=0$.

\textit{2) Residential-load-aware prediction:}
We introduce MPC-FB, which uses the previous interval's loading correction $\Delta S_k$ to account for reconstructed residential charging and AC feeder losses. Let $E_0^{\mathrm{res}}$ (kWh) denote the baseline daily residential charging requirement without DWC, estimated day-ahead from arrival demand. The controller assumes $E_0^{\mathrm{res}}$, $\sigma$, and $h_k$ are known, with estimation errors examined in Sec.~\ref{sec:results}. Using these estimates and cumulative metered DWC energy, we reconstruct:
\begin{align}
\tilde p_{k-1}^{\mathrm{res}}
&=\frac{10^{-3}\sigma h_{k-1}}{\eta}
\pos{E_0^{\mathrm{res}}-\eta E_{k-1}^{\mathrm{DWC}}},
\label{eq:rec-ph}\\
\tilde S_{k-1}
&=N\,\mathcal F_{\mathrm{AC}}(\tilde{\mathbf p}_{k-1},
\tilde{\mathbf q}_{k-1}),
\label{eq:rec-s}\\
\Delta S_k
&=\tilde S_{k-1}
-\hat S_{k-1}(\bar p_{k-1}^{\mathrm{DWC}}).
\label{eq:rec}
\end{align}

Here, $\tilde{\mathbf p}_{k-1},\tilde{\mathbf q}_{k-1}\in\mathbb R^J$ are reconstructed bus demands from \eqref{eq:busp}--\eqref{eq:busq}, using $\tilde p_{k-1}^{\mathrm{res}}$. We evaluate $\hat S_{k-1}$ from \eqref{eq:pred} using the preceding interval's average DWC power, initializing $\Delta S_1=0$. Since \eqref{eq:rec-ph} subtracts only DWC energy delivered to date, it conservatively overestimates residential demand relative to the retrospective daily allocation. We then apply $\Delta S_k$ throughout the subsequent prediction horizon. This model-based correction does not use measured residential demand or guarantee feeder loading compliance.

\textit{3) Comparison policies:}
We compare both controllers against unpriced, time-of-day (TOD), and feedback (FB) pricing. The unpriced policy sets $\mathbf u_k=\mathbf 0$. TOD uses a fixed schedule obtained by averaging prices over 64 simulated days from an earlier MPC formulation with a system-level energy-stress price, applied to the 2006 eastbound I-210 corridor. FB retains the TOD toll and adjusts the CPA using gain $\gamma$ (USD/kWh) and normalized loading deviation $(\tilde S_{k-1}-\theta S^{\max})/S^{\max}$, subject to the TOD CPA floor and price bounds, where $\theta$ is the loading threshold fraction. Three forecast-based variants assess residential demand estimation and feeder losses: MPC-OC uses a day-ahead residential forecast, MPC-OCu updates it using cumulative DWC delivery, and MPC-OCuL additionally includes AC losses. MPC-FB-P penalizes forecast daily peak feeder power with weight $w_P$ (USD/kW), assuming no additional DWC demand beyond the prediction horizon.

\subsection{Performance Metrics}\label{sec:outcomes}

We evaluate the pricing strategies based on electrical, transportation, and economic performance. To quantify feeder loading violations, we define the daily overload as:
\begin{equation}\label{eq:OL}
S^{\mathrm{over}}=\Delta_k\sum_{k\in\mathcal K}
\pos{S_k-S^{\max}},
\end{equation}
where $S^{\mathrm{over}}$ (MVA$\cdot$h) measures the cumulative magnitude and duration of loading exceedances. We report the number of days with $S^{\mathrm{over}}>0$, along with daily maximum and evening (19:00--21:00) peak feeder loading relative to a no-DWC reference with baseline residential charging.

For transportation performance, we report canceled trips and aggregate traffic delay $D$ (veh$\cdot$h). We estimate delay as the additional travel time relative to free-flow conditions, calculated from accumulated vehicle-hours in each roadway cell minus the corresponding free-flow traversal time of departing EVs, with negative differences truncated to zero. We also include waiting time in entry queues. Canceled trips are excluded from this delay metric. We calculate daily operator net revenue $R$ (USD) from roadway tolls and electricity sales, accounting for electricity procurement and operating costs:
\begin{equation}\label{eq:R}
R\!=\!\sum_{t\in\mathcal T}\!
\left[\tau_t^{\mathrm{bill}}\!+\!\!
\left(\!\!\pi^{\mathrm{DWC}}\!+\!\mu_{k(t)}
\!-\!\frac{c_t^{\mathrm{wh}}}{\eta}\!-\!c^{\mathrm{om}}\right)
\!E_t^{\mathrm{del}}\right],
\end{equation}
where $\tau_t^{\mathrm{bill}}$ (USD) is the toll revenue collected at $t$, $c_t^{\mathrm{wh}}$ (USD/grid kWh) is the wholesale electricity price, and $c^{\mathrm{om}}$ (USD/battery kWh) is the operating cost per unit of delivered energy. Capital and electricity delivery costs are excluded.

We evaluate driver savings relative to a no-DWC scenario in which all charging requirements are met at home, with baseline expenditure $\pi^{\mathrm{res}}E_0^{\mathrm{res}}$. Let $b$ (USD) denote total daily driver expenditure on tolls, DWC charging, and residential charging. Driver savings are $\pi^{\mathrm{res}}E_0^{\mathrm{res}}-b$. We further define partial social benefit as $B=R-b-w_DD$ (USD) and report its change relative to the unpriced DWC scenario, $\Delta B=B-B^{\mathrm{unpriced}}$. This metric excludes feeder overload costs and the welfare impacts of canceled trips.

\section{Case Study}\label{sec:case}

\subsection{Simulation Setup and Assumptions}\label{sec:setup}

\begin{table}
\centering
\caption{Simulation and controller parameters}
\label{tab:parameters}
\scriptsize
\setlength{\tabcolsep}{3pt}
\renewcommand{\arraystretch}{1.05}

\begin{tabular}{@{}llll@{}}
\toprule
Parameter & Value & Parameter & Value \\
\midrule
$\Delta t$ & 10 s & $\Delta_k$ & 15 min \\
$I$ & 40 & $J$ & 33 \\
EV shares (\%) & 93.49/1.82/4.69 & $p^{\mathrm{veh}}$ & 190 kW \\
$e_c^{\mathrm{cap}}$ (kWh) & 36.5/97/226 & $\eta$ & 0.90 \\
Traction (kWh/km) & 0.14/0.66/0.78 & Aux. (kW) & 2/2/7.5 \\
Initial energy (\%) & 30/50/70 & Targets (\%) & 60/90 \\
$\kappa$ & 0.50 & $\xi$ & 1.00 \\
$j^{\mathrm{DWC}}$ & 6 & $\cos\varphi$ & 0.98 \\
Source voltage & 1.05 p.u. & $\sigma$ & 0.25 \\
Residential profile & Mean 20:30, SD 4 h & $\phi_0$ & 0.25 \\
$\pi^{\mathrm{DWC}}$ & 0.22 USD/kWh & $\pi^{\mathrm{res}}$ & 0.45 USD/kWh \\
$\tau$ bounds (USD) & $[0,12]$ & $\mu$ bounds (USD/kWh) & $[-0.10,0.50]$ \\
$\varepsilon$ & 0.18 USD$^{-1}$ & $\beta$ & 2 kWh/USD \\
$\chi$ & 0.85 & Horizon & 45 min \\
$\delta_\tau$ & 1 USD & $\delta_\mu$ & 0.05 USD/kWh \\
$w_R$ & 0.10 & $w_D$ & 24 \\
$w_S$ & 100 & $w_L$ & 100 \\
$w_P$ & 1 & $c^{\mathrm{om}}$ & 0.010 USD/kWh \\
$\gamma$ & 2 USD/kWh & $\theta$ & 0.995 \\
\bottomrule
\end{tabular}
\end{table}

We evaluate the framework on the 22.6-km I-210 West corridor using CTMSIM \cite{kurzhanskiy2009, ghose2025dwptgrid}, with 40 cells per lane group and capacities adjusted by lane count. We use 78 days of traffic records from 2024 (56 weekdays) from 12 calendar-week blocks, excluding six days with missing detector data or unsuccessful calibration. Daily demand profiles are scaled using hourly Caltrans PeMS measurements~\cite{caltranspems}.

We assume all traffic consists of EVs, classified as light, medium, or heavy, with heterogeneous battery capacities, energy consumption, and initial charging states. Each feeder follows the 33-bus, 12.66-kV radial benchmark~\cite{baran1989}.
We generate base loads and wholesale electricity prices through seasonal resampling of 2024 demand in California~\cite{eia930} and PG\&E-zone day-ahead prices~\cite{caisooasis}, using monthly and day-type median profiles, common donor-day residuals, and seeded amplitude perturbations. Residential charging follows a wrapped normal plug-in time distribution~\cite{powell2022}, with a mean of 20:30 and standard deviation of 4 h. Table~\ref{tab:parameters} summarizes the simulation and controller parameters. The assumed 190-kW charging limit for all EV classes is optimistic for light-duty EVs.

The TOD benchmark uses fixed tolls of 1.05, 3.52, and 0.33 USD and CPAs of $-0.05$, 0.105, and $-0.08$ USD/kWh during 07:00--17:00, 17:00--23:00, and 23:00--07:00, respectively.

\subsection{Evaluation Scenarios}\label{sec:scenarios}

We consider two feeder configurations. The \emph{voltage-feasible case} uses $N=128$, the smallest power of two maintaining bus voltages above 0.95 p.u. without DWC, with $S^{\max}=680.5$ MVA set by the no-DWC peak. The \emph{stress case} uses fewer feeders ($N=32$) and a relaxed 0.85-p.u. voltage threshold under no-DWC and unpriced conditions, representing a more constrained network. Its limit, $S^{\max}=246.4$ MVA, is obtained by scaling unpriced DWC and residential loads to 80\%.

We evaluate all policies over 78 days using retrospective AC power flow. Sensitivity tests vary residential demand estimates, feeder share $\sigma$, charging timing, correction lag, prediction noise, and individual pricing controls. Four additional scenarios consider evening charging ($\sigma=0.5,1.0$) and overnight charging ($\sigma=0.25,0.5$), with $N$ and $S^{\max}$ recalculated.

\subsection{Results and Discussion}\label{sec:results}

\textit{1) Comparative performance:}
Under the voltage-feasible configuration, unpriced, TOD, FB, and MPC exceed the feeder loading limit on 76, 34, 34, and 52 of 78 days, respectively. MPC-FB eliminates this with only 0.6\% lower revenue than MPC. Similar results hold under the stress configuration, where MPC-FB eliminates MPC's 50 overloaded days at a 0.9\% revenue cost (Table~\ref{tab:results}). These results suggest that explicitly accounting for residential charging can substantially improve feeder loading management without sacrificing revenue.

\begin{table}
\centering
\caption{Electrical, economic, and transportation performance over 78 simulation days under the stress configuration ($N=32$)}
\label{tab:results}
\scriptsize
\setlength{\tabcolsep}{3pt}
\renewcommand{\arraystretch}{1.05}

\begin{tabular}{@{}lrrrr@{}}
\toprule
\multicolumn{5}{c}{\textit{Electrical performance}}\\
\midrule
Policy & Days over & Overload & Daily peak & Evening peak\\
       & limit & (MVA$\cdot$h) & (\%) & (\%)\\
\midrule
Unpriced & 77 & 32.4 & +11.1 & +11.2\\
TOD & 49 & 4.1 & +6.2 & +0.3\\
FB & 49 & 1.6 & +5.6 & $-0.2$\\
MPC & 50 & 11.2 & +7.5 & +7.5\\
MPC-FB & 0 & 0.0 & +4.2 & +4.2\\
MPC-FB-P & 0 & 0.0 & $-3.7$ & $-3.6$\\
\bottomrule
\end{tabular}

\vspace{5pt}

\begin{tabular}{@{}lrrrrr@{}}
\toprule
\multicolumn{6}{c}{\textit{Economic and transportation performance}}\\
\midrule
Policy & Revenue & Delay & Canceled & Driver & Social\\
       & (kUSD) & (veh$\cdot$h) & trips & savings & benefit\\
       &        &               &       & vs.\ no lane & vs.\ unpriced\\
\midrule
Unpriced & 114.0 & 11,620 & 0 & +161.6 & 0\\
TOD & 124.8 & 12,210 & 830 & +116.9 & $-48.0$\\
FB & 125.0 & 12,210 & 840 & +115.6 & $-49.1$\\
MPC & 234.5 & 9,660 & 1,990 & +0.4 & +6.4\\
MPC-FB & 232.5 & 9,660 & 1,970 & $-3.6$ & +0.2\\
MPC-FB-P & 213.1 & 9,730 & 1,840 & +0.4 & $-16.7$\\
\bottomrule
\end{tabular}

\vspace{3pt}
\parbox{\columnwidth}{\scriptsize

\textit{Note:} Daily values are 78-day averages. Peak changes are relative to no-DWC; evening denotes 19:00--21:00. Monetary metrics are in $10^3$ USD/day.

}
\end{table}

\textit{2) Role of residential charging:}
The improved performance of MPC-FB arises from correcting for residential demand and feeder losses omitted by lane-only prediction. During 17:00--23:00 under MPC, residential charging averages 63.8 MW on the feeders (61.1 MW under the unpriced lane), exceeding the 28.5-MW DWC demand. When MPC exceeds the stress limit, actual loading exceeds its prediction by 75.6 MVA on average, comprising 77.6 MVA of baseline residential demand, offset by 15.1 MVA displaced through DWC delivery, and supplemented by 13.1 MVA of incremental network losses. This highlights why modeling the interaction between roadway and residential charging is essential for accurate feeder loading predictions.

\begin{figure}
\centering
\includegraphics[width=\columnwidth]{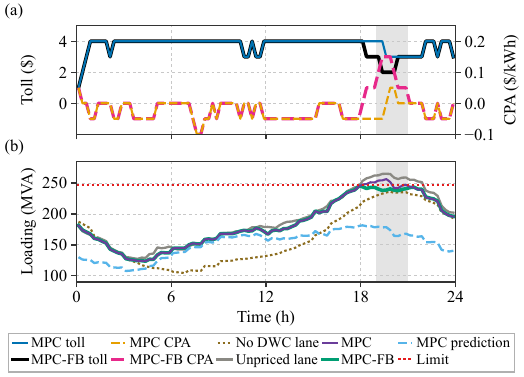}
\caption{Representative day (July 11, 2024) illustrating toll and charging-price decisions and their impact on feeder loading under the stress configuration. The shaded region indicates the evening peak period (19:00--21:00).}
\label{fig:day}
\end{figure}

Fig.~\ref{fig:day} illustrates the pricing response on July 11, 2024. As corrected loading approaches the limit around 18:00, MPC-FB raises the CPA while lowering the entry toll earlier than MPC, discouraging charging without proportionally reducing lane participation. This prevents the overload observed under lane-only MPC, demonstrating the benefit of independently controlling roadway access and energy consumption.

\textit{3) Robustness and pricing sensitivity:}
Forecast-based comparisons reveal the importance of both residential demand estimation and incremental feeder losses. MPC-OC avoids violations by overestimating evening residential charging by 22\%, at a 1.6\% revenue cost relative to MPC. However, MPC-OCu, which updates demand forecasts but neglects incremental losses, exceeds the limit on 50 days, while MPC-OCuL eliminates these violations. MPC-FB remains sensitive to estimation errors: underestimating residential demand by 20\% or feeder share by 25\%, or shifting charging one hour later results in 50 overloaded days. Finally, toll-only control leaves 25 overloaded days, while CPA-only control leaves seven but reduces revenue by 44\% and increases delay by 16.4\%. These results highlight the need for accurate electrical modeling and coordinated pricing to balance feeder compliance with economic and traffic performance.

\begin{figure}[t]
\centering
\includegraphics[width=\columnwidth]{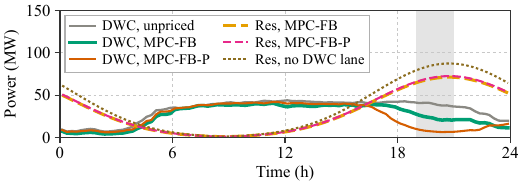}
\caption{Average daily DWC and residential charging power profiles across 78 days under different pricing strategies in the stress configuration. The shaded region indicates the evening peak period (19:00--21:00).}
\label{fig:shift}
\end{figure}

\textit{4) Peak demand and economic implications:}
Avoiding feeder overloads does not necessarily reduce peak demand. Under the stress configuration, MPC-FB increases the daily peak by 4.2\%, while MPC-FB-P reduces it by 3.7\% relative to no-DWC, at an 8.3\% revenue cost (Fig.~\ref{fig:shift}). Across four alternative residential charging scenarios, neither MPC nor MPC-FB exceeds the limit, as residential demand determines the peak. Thus, the value of residential-load-aware control depends on the relative contributions of DWC and residential charging. Under the assumed driver responses \eqref{eq:choice}--\eqref{eq:optin}, MPC and MPC-FB more than double operator revenue relative to unpriced operation and earn over 85\% more than TOD and FB, but reduce driver savings. The 17\% reduction in modeled delay also excludes nearly 2,000 canceled trips daily, leaving partial social benefit largely unchanged.

\section{Conclusion}\label{sec:conclusion}

We develop a grid-aware DWC pricing framework that jointly optimizes tolls and charging prices while capturing their effects on residential charging and feeder losses. Results demonstrate that accounting for displaced residential demand improves feeder loading management with minimal revenue loss, while coordinated pricing balances roadway utilization and grid constraints. Future work will calibrate driver responses and incorporate more realistic residential charging schedules. We will also develop pricing controllers that account for uncertainty in residential charging demand and its timing.

% \balance
\bibliographystyle{IEEEtran}
\bibliography{refs}

\end{document}